\documentstyle[letter,epsf]{ptptex}

\title{
Is Angular Momentum in an Accretion Disk Transported Inwards?
}

\author{
Eiji {\sc Hayashi}
and Takuya {\sc Matsuda}
}

\inst{
Department of Earth and Planetary Sciences, Kobe University,\\
 Kobe 657-8501, Japan
}

\recdate{October 6, 2000
}

\abst{
The derivation of the viscosity formula of accretion disks 
appearing in a number of textbooks is based on a mean 
free path theory of gas particles.  However, this procedure, 
when followed precisely, leads to the incorrect conclusion that
 the angular momentum of an accretion disk flows from the outer part 
to the inner part.}

\begin{document}

\maketitle

{\it Introduction\/} 
The standard model of accretion disks, 
also called the \(\alpha\) disk model, proposed 
by Shakura and Sunyaev\cite{rf:1} states 
that the angular momentum is transported from the inner part 
of the disk to the outer part, due to the action of 
some kind of viscosity. The Reynolds number of the flow in 
accretion disks is estimated to be as high as \( 10^{11} \), 
which means, according to the accepted hydrodynamical wisdom, 
that the flow is turbulent (Spruit\cite{rf:2}).  It has been 
assumed that in calculating the viscous stress due to the 
turbulence one can employ the usual molecular viscosity 
formula.

There are many textbooks and reviews about the formation of accretion 
disks and the mechanism involved in the transport of 
angular momentum (e.g., Pringle,\cite{rf:3}, Frank, King and 
Raine,\cite{rf:4}
 Spruit,\cite{rf:2} Hartmann,\cite{rf:5} and Kato, Fukue and
Mineshige,\cite{rf:6}).
Some of these textbooks,\cite{rf:4,rf:5} however, give elementary explanations 
about the phenomenon of 
angular momentum transport based on physical arguments, i.e. 
a mean free path theory, rather 
than with mathematical rigor. 
This is perhaps because the authors of such textbooks intend 
to make the explanation easier for beginners to understand.
We find that these explanations in fact do not lead to the 
results intended by the authors but to the incorrect conclusion 
that {\em the angular momentum in an accretion disk is transported 
inwards\/}.
  To the contrary, the angular momentum must be transported outwards, 
or accretion will not occur.  There are, however, 
some other textbooks and reviews that 
give rigorous mathematical derivations of formulas describing 
the angular momentum transport, starting from the formula 
for the molecular viscosity (e.g., Spruit,\cite{rf:2} 
Kato et al.\cite{rf:6}).

The above mentioned ``elementary explanations'' based on the mean free path 
theory, however, seemed very plausible 
to us. We therefore began to doubt if the ``molecular viscosity 
formula'' itself is applicable to gas in an accretion disk 
rotating with Keplerian motion in the gravitational field of 
a central compact star.  Eventually, we found that these 
``elementary explanations'' lead to the incorrect conclusion and 
that the ``molecular viscosity formula'' is basically correct.\\

{\it Is mean free path theory applicable to the present problem?\/}
In the standard disk model, the equation for angular momentum 
conservation is
$$     \frac{\partial}{\partial t}(R \Sigma \Omega R^2) 
+ \frac{\partial}{\partial R}
(R \Sigma v_r \Omega R^2) = \frac{\partial}{\partial R}
\left (\Sigma \nu R^3 \frac{\partial
\Omega}{\partial R} \right ),  \eqno{(1)} $$
where 
\begin{eqnarray}
   R & = & \mbox{distance from the center of the central star},
\, \Sigma = \mbox{surface density}, \nonumber \\
   \Omega & = & v_{\phi}/R,\,\mbox{and} \,\,
      \nu  =  \mbox{kinematic viscosity.} \nonumber 
\end{eqnarray}

How can this equation be derived? It can be obtained using 
the formula of molecular viscosity for incompressible gas 
(e.g., Spruit\cite{rf:2})

$$    \sigma_{ik} = \mu \left ( \frac{\partial v_i}{\partial x_k}
 + \frac{\partial v_k}{\partial x_i} \right ),   \eqno{(2)}
$$
where  \( \sigma_{ik}\) is the stress tensor, 
\( \mu \) is the viscosity coefficient, and \(v_i \) and \(v_k \) 
are averages of the molecular velocity 
in the {\it i}- and {\it k}-direction, respectively.  The viscous force is

$$    F_i = \frac{\partial \sigma_{ik}}
{\partial x_k} = \frac{\partial \mu}{\partial x_k}
 \left ( \frac{\partial v_i}{\partial x_k}
 + \frac{\partial v_k}{\partial x_i} \right ) + \mu \triangle v_i.
 \eqno{(3)} $$
A problem emerged when we were learning with a rather elementary textbook,
which gives explanations for beginners. 
Since the situation is best described in the textbook, we make a 
rather lengthy quotation from
{\it Accretion Processes in Star Formation\/}, by Hartmann,\cite{rf:5}
p. 30 (the latter part of \S5.1):\\

\begin{quotation}
Following Frank et al. (1992), we can calculate the magnitude of 
the angular momentum transfer in terms of a {\it kinematic 
viscosity\/}.  The basic picture is one in which turbulent 
elements of the gas moving at a typical random velocity \( w \)
travel a mean free path \( \lambda \) before mixing with 
other material.  Thus, the net torques or angular momentum transfer 
at cylindrical radius \( R \) (Figure \(5.1)\) will be produced 
(schematically) by the differing angular momenta of two streams of 
material; one from material originating at \( R - \lambda/2 \) 
 and moving outward across \( R\) to mix with annular material
 centered at \(R + \lambda/2 \); and the other starting at 
\( R + \lambda/2\) and moving inward across \( R\) to mix 
with the inner annulus at \(R - \lambda/2 \).

\hspace{1mm}
In this kinematic viscosity model, no net angular momentum is transported 
unless there is shearing orbital motion, \(d \Omega/dR \equiv 
\Omega' \ne 0 \). The net angular momentum fluxes can be calculated in 
the following way. Material originating at \( R - \lambda/2 \) has an 
angular momentum of
$$
 J_{in} = (R - \lambda/2)^2 \Omega(R- \lambda/2)
 = (R - \lambda/2)^2 [\Omega(R) - (\lambda/2)(d \Omega/dR)],
\eqno{(5.9)}  $$
where we have approximated the difference in the angular velocities 
in terms of the first derivative with respect to \( R \).  A 
similar expression with the negative values changed to positive
 applies to the material at \( R + \lambda/2 \).  The inner 
material diffuses outwards at velocity \( w \) and the 
outer material diffuses inward at \( w \) across \( R \). (The {\it net\/}
 inward motion of material in the accretion disk is assumed to be
small in comparison with the turbulent velocity \( w \)).
For simplicity we integrate or average the disk structure 
in the direction perpendicular to the disk plane; then the net 
outward transfer of angular momentum across \( R \) per unit 
length for a disk with mass density 
per unit area \( \Sigma \) is
$$  \Sigma w [ (R - \lambda/2)^2( - \lambda/ 2)d \Omega /dR 
  -(R + \lambda/2)^2(\lambda/2)d\Omega/dR ]  
=  - \Sigma w \lambda R^2 d \Omega /dR,  
\eqno{(5.10)} $$
where we have assumed that \( \lambda \)
is a short distance compared with the scale over which \(\Omega \)
varies significantly. With this result, the total angular momentum
flux outward across \( R \), i.e. the torque of the inner annulus on
the outer annulus, can be written as
$$ g = -2 \pi R \Sigma \nu_v R^2 d\Omega/dR,  \eqno{(5.11)} $$
where the viscosity \(\nu_v \) is
$$   \nu_v = \lambda w.  \eqno{(5.12)}  $$

Note that a negative gradient of angular velocity (i.e.\(\Omega \) 
decreasing outward) leads to a positive outward flux 
of angular momentum, as predicted by the qualitative picture 
of friction between neighboring annuli discussed above.\\

\end{quotation}

Hereafter, we explain why the above derivation of (5.11) is not 
correct. In Eq. (5.9), \( (R - \lambda/2) \) in the original textbook 
should be \( (R - \lambda/2)^2 \), and is so corrected in the 
present text.  From (5.9) and a similar expression for \(J_{out} \), 
we have
\begin{eqnarray}
 J_{in} - J_{out}
 & = & \Omega(R)[ (R - \lambda/2)^2-(R + \lambda/2)^2] \nonumber \\
 & - & (\lambda/2) d\Omega/dR[ (R - \lambda/2)^2 + (R + \lambda/2)^2] 
\nonumber \\
 & = & \Omega(R)(-2R\lambda)-(\lambda/2) d\Omega/dR (2R^2) 
 = -\lambda d(R^2\Omega)/dR. \nonumber 
\,\,\,\,\quad \quad \qquad \qquad (4)
\end{eqnarray}
Then, this gives, not Eq. (5.11), but

$$  g = - 2 \pi R \Sigma \nu_v d(R^2 \Omega)/dR. \eqno{(5)} $$

The difference between Eqs. (5.11) and 
(5) is that the factor \( R^2 \) is operated on by 
the \( d/dR \) operator in Eq. (5) 
and not operated on by it in  
Eq. (5.11).
If Eq. (5.11) is correct, the angular momentum 
flows in such a direction as to realize \( \partial \Omega/
\partial R = 0 \), that is, to produce rigid body rotation.
 In that case, the angular momentum flows outwards, which is 
what the standard theory predicts.

  Attention 
should be paid here to the following.  That is, with a Keplerian 
disk, we have \(\Omega \propto R^{-3/2} \), \(v_{\phi} = R\Omega 
\propto R^{-1/2}\) and \( L = Rv_{\phi} 
\propto R^{1/2} \), where \(v_{\phi} \) is the circumpherential 
velocity and \( L\) is the angular momentum.  An accretion disk 
has the characteristic features that \(\Omega\) and \( v_{\phi} \) 
decrease with increasing \(R\) and that \(L \), in constrast, 
increases with increasing \(R\).
(See Fig. 1 to better understand the above description.)

If, in contrast, Eq. (5) holds, 
the angular momentum flows in the direction to realize
 \( d (R^2 \Omega)/
d R = 0 \), that is, to a constant angular momentum.  
In this case the 
\begin{figure}[h]
\epsfxsize = 11 cm
\centerline{\epsfbox{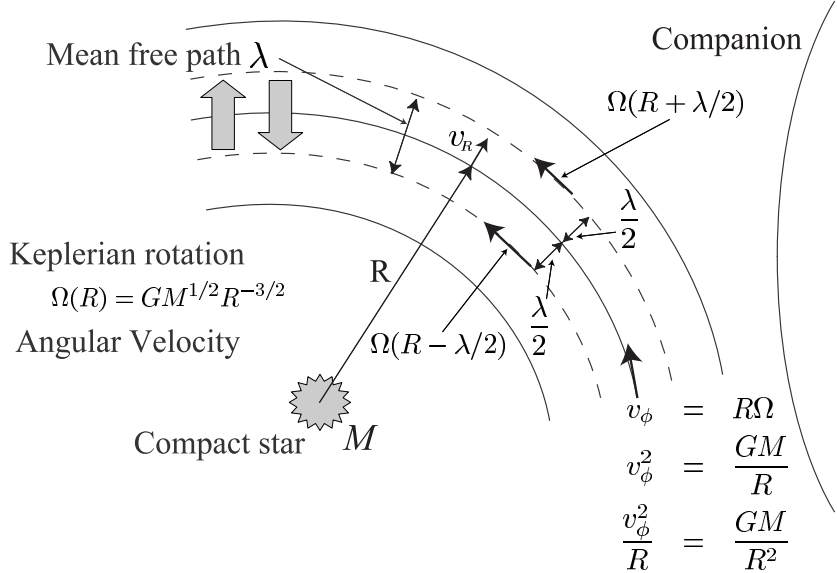}}
\caption{Angular momentum transport between adjacent annuli
in a thin accretion disk.  The angular velocity \( \Omega \) and 
circumferential velocity \(v_\phi = R \Omega \) decrease with 
increasing distance \(R \).  The angular momentum \( R^2 \Omega \), 
however, increases 
with increasing \(R \). The big wide arrows indicate 
which way the angular momentum is 
transported, inwards or outwards.}
\end{figure}
angular momentum flows inward. 
This means that even an originally disk-shaped gas will 
end up as a ring.  We conclude here that the correct manipulation 
of Hartmann's procedure leads to the conclusion that he did not 
intend.\\

There is another famous textbook entitled {\it Accretion Power 
in Astrophysics\/} written by Frank, King and 
Raine.\cite{rf:4}  There, the authors give a similar elementary 
argument, with a similar confusion, while referring to a similar figure, 
as follows (on pages 58 and 59 of the second
edition)\footnote{Interestingly, 
this description has been revised and 
differs from the corresponding one in the first edition
(1985). 
There, 
the ``correct'' conclusion is derived from the incorrect
 assumption that 
a gas element in a disk would, in chaotic motion, move radially 
a distance of order \(\lambda \) away, while carrying its {\em momentum
\/}(not its angular momentum). Since the rotating gas element 
should carry its angular momentum on that occasion, this assumption 
is wrong.  This may have been the 
reason for the revision in the second edition.}: \\
\begin{quotation}
... As these elements of fluid are exchanged, they carry slightly 
different amounts of angular momentum: elements such as \(A\) will 
{\it on average\/} carry an angular momentum corresponding to 
the location \(R - \lambda/2 \), while elements such as \(B\) will 
be representative of a radial location \( R + \lambda/2 \)...
\\
\end{quotation}

If the authors had proceeded in this way correctly, they 
should have obtained the same result as from Hartmann's 
``correct'' (i.e. corrected in the present paper) manipulation.  
They, instead, 
continue as follows:

\begin{quotation}
... As seen by an observer corotating with the fluid at \(P\)
 [at R]
with angular
velocity \(\Omega(R)\), the fluid at \(R- \lambda/2\) will appear 
to move with velocity \( (R-\lambda/2)\Omega(R-\lambda/2)+ \Omega(R)
\lambda/2 \).  Thus the average angular momentum flux per unit arc 
length through \( R = \) const in the outward direction is 
$$
    \rho \tilde{v} H(R - \lambda/2)[(R - \lambda/2)\Omega(R-\lambda/2)
+\Omega(R)\lambda/2]
 \nonumber \eqno{(a)}
$$
[where \(\tilde{v} \) is the speed of gas elements in a chaotic
 motion,
 and the formula numbers (a) and (b) are temporarily put by the
present authors].
An analogous expression, changing the sign of \(\lambda \), gives the 
average inward angular momentum flux per unit arc length.  The 
torque exerted on the outer ring by the inner ring is given by 
the {\it net \/} outward angular momentum flux.  Since the mass flux
due to chaotic motions is the same in both directions, one obtains 
to first order in \(\lambda\) the torque per unit arc length as
$$
   - \rho\tilde{v} H \lambda R^2 \Omega',   \eqno{(b)}
$$
where \( \Omega' = d\Omega/dr \) [and \(H\) is the half thickness of 
the disk]...
\\
\end{quotation}

  Although the meaning of (a) is not clear and is discussed 
below, let us accept (a)
 for the moment.  We denote the formula (a) by \( L_{in}\) and 
make a linear approximation, leading to:
\begin{eqnarray}
L_{in} & = & \rho\tilde{v} H(R - \lambda/2)[(R
       -\lambda/2)(\Omega(R)-(\lambda/2)\Omega(R)')+\Omega(R)\lambda/2]
       \nonumber \\
       & = & \rho\tilde{v} H (R - \lambda/2)[R\Omega - 
R(\lambda/2)\Omega']. \nonumber \qquad \qquad \qquad \qquad 
\qquad \qquad \qquad \qquad \qquad \,\, (6)
\end{eqnarray}
Similarly, 
$$
L_{out}  =   \rho\tilde{v} H (R + \lambda/2)[R\Omega + R(\lambda/2)\Omega'].
 \eqno{(7)}  $$
Then, the torque should not be, \(- \rho\tilde{v} H \lambda R^2
\Omega' \), 
but rather
\begin{eqnarray}
L_{in} - L_{out} & = &
  \rho\tilde{v} H[(R - \lambda/2)(R\Omega-R(\lambda/2)\Omega')
-(R+\lambda/2)(R\Omega +R(\lambda/2)\Omega')] \nonumber\\                 
& = &  -\lambda\rho\tilde{v}HR(\Omega + R\Omega') 
 =  -\nu\Sigma R d(R\Omega)/dR. \nonumber \qquad \qquad \qquad \qquad
\qquad \quad \, (8)
\end{eqnarray}
If this is true, the angular momentum flows in a direction to 
realize \(d (R\Omega)/d R = 0 \), which implies 
a constant azimuthal velocity and is unrealistic in a rotating gas, 
although in this case the angular momentum flows outward.

Now we return to the problem of (a).  When seen from \(P\),
the fluid at \(R- \lambda/2\) should appear 
to move with velocity \( (R-\lambda/2)\Omega(R-\lambda/2) 
\underline{- R\Omega(R)} + \Omega(R)
\lambda/2 \), which completes the transformation from the inertial frame 
to the corotational frame with the origin at \(P\).
  Eq. (a) then should be

$$    \rho \tilde{v} H(R - \lambda/2)[(R - \lambda/2)\Omega(R-\lambda/2)
- R\Omega(R) +\Omega(R)\lambda/2],
 \eqno{(a)'} $$
so that    
$$ L_{in} = \rho\tilde{v} H[(R - \lambda/2)(-R(\lambda/2)\Omega')
= - \rho\tilde{v} H R^2 (\lambda/2)\Omega', \eqno{(b)'} $$
and
$$ L_{in}-L_{out} = 
- \rho\tilde{v} H R^2 \lambda \Omega'.  \eqno{(8)'} $$

This is the correct result.  We then, however, notice that the procedure
to obtain \((8)'\) is simply that to obtain the 
well-known viscosity formula 
(2), thus being no longer based on the mean free path theory.\\ 
 
{\it Discussion\/}
Having pointed out the confusion in the derivations given in 
the above cited textbooks, we now clarify how naturally the 
wrong conclusion comes 
from application of the mean free path theory
(for discussion of 
this theory see, for example, Vincenti and Kruger\cite{rf:7}
 Chap. 1, \S 4.) to a rotating gas in an accretion disk.

A gas element rotating in a Keplerian orbit possesses an angular momentum, 
linear momentum and angular velocity.
  The element is then assumed to travel across \(R\)
over a distance of \(\lambda \), the mean free path.  If we denote the 
quantity to be transported then by the element as \(Q \), which is
the angular momentum, 
linear momentum or angular velocity, then, by linear approximation,
$$    Q(R + \lambda) - Q(R) = \lambda dQ(R)/dR.
\eqno{(9)} $$
Thus, if we want to obtain the correct result, i.e. to 
introduce the angular velocity alone in \(d/d R\), 
we have to make the impossible assumption that \(Q\) is 
the angular velocity.
This argument clearly shows the failure of the application of 
the mean free path theory.

Do we then have to use the molecular viscosity formula to 
consider the present problem? This is the next question.
In a strict sense, the molecular viscosity formula (2) is derived for 
a non-rotating gas.  A viscosity formula for a rotating gas should be 
derived using the Boltzmann equation with the Coriolis force taken
into account.  For a rigidly rotating gas, this was 
formulated by Chapman and Cowling,\cite{rf:8} who considered the 
effect of the Lorentz force rather than the Coriolis force, with the
result that 
the viscosity coefficient along the direction parallel 
to the rotation axis is the same as that in a non-rotating gas, while 
that perpendicular to the axis is suppressed in a manner that depends 
on the Knudsen 
number.
The viscosity formula for a Keplerian rotating gas is much more 
complicated, and an exact formula cannot be obtained.  Approximate 
formulas obtained by a few authors are summarized in the textbook 
of Fridman 
and Gorkavyi.\cite{rf:9}


\vspace*{3mm}
The authors thank Professor S. Mineshige for his valuable comments and advice.



\end{document}